\documentclass[entropy,article,accept,oneauthor,pdftex,12pt,a4paper]{mdpi}
\pdfoutput=1  
\lastpage{x}
\doinum{10.3390/------}
\pubvolume{17}
\pubyear{2015}
\externaleditor{Academic Editors: Fr\'ed\'eric Barbaresco and Ali Mohammad-Djafari}
\history{Received: 3 April 2015 / Accepted: 25 June 2015 / Published: xx }
\usepackage{graphicx}
\usepackage{subfigure}
\usepackage{amssymb}
\usepackage{booktabs,multirow}
\usepackage{enumerate}
\usepackage{bm}
\usepackage{soul}
\usepackage{upgreek}
\usepackage{booktabs}
\newcommand{\up}[1]{\mathrm{#1}}
\newcommand{\ud}{\up{d}}

\newcommand{\p}{\ensuremath{\partial}}
\newcommand{\dt}[1]{\ensuremath{\dot{#1}}}
\newcommand{\ddt}[1]{\ensuremath{\ddot{#1}}}
\newcommand{\G}[3]{\ensuremath{\Gamma^{#1}_{\phantom{#1}#2#3}}}

\Title{A New Robust Regression Method Based on Minimization of Geodesic Distances on a Probabilistic Manifold: Application to Power Laws $^\dagger$}
\Author{Geert Verdoolaege $^{1,2,}$*}
\address{%
$^{1}$ Department of Applied Physics, Ghent University, Sint-Pietersnieuwstraat 41, B-9000 Ghent, Belgium; E-Mail: geert.verdoolaege@ugent.be; Tel.: +32-9-264-95-91; Fax: +32-9-264-41-98\\
$^{2}$ Laboratory for Plasma Physics---Royal Military Academy (LPP-ERM/KMS), Avenue de la Renaissancelaan 30, B-1000 Brussels, Belgium}
\abstract{In regression analysis for deriving scaling laws that occur in various scientific disciplines, usually standard regression methods have been applied, of which ordinary least squares (OLS) is the most popular. In many situations, the assumptions underlying OLS are not fulfilled, and several other approaches have been proposed. However, most techniques address only part of the shortcomings of OLS. We here discuss a new and more general regression method, which we call {geodesic least squares regression} (GLS). The method is based on minimization of the Rao geodesic distance on a probabilistic manifold. For~the case of a power law, we demonstrate the robustness of the method on synthetic data in the presence of significant uncertainty on both the data and the regression model. We~then show good performance of the method in an application to a scaling law in magnetic confinement~fusion.}

\keyword{regression analysis; information geometry; geodesic distance; scaling laws; nuclear fusion}
\conference{the 34th International Workshop on Bayesian Inference and Maximum Entropy Methods in Science and Engineering (MaxEnt 2014), \mbox{21--26 September 2014}, Amboise, France}

\begin{document}
%
\newpage 
\section{Introduction}

Regression analysis is an essential instrument for data analysis in numerous branches of science. It is used for investigating deterministic relations between variables, for model building and for prediction by extrapolation to a previously unseen range of the involved variables. In this paper, we focus on regression analysis applied to the estimation of scaling laws. In various scientific disciplines, such as astronomy, biology, ecology and geology, scaling laws are used to characterize the underlying mechanisms at work in the respective complex systems under study. In general, a scaling law describes how a quantity of interest $y$ {scales} when changing other quantities $x_1, x_2,
\ldots, x_P$, on which it depends. Scaling laws are often expressed in terms of a {power law}:
\begin{equation} \label{eq:power_law}
 y = \upbeta_0x_1^{\upbeta_1}x_2^{\upbeta_2}\ldots x_P^{\upbeta_P}.
\end{equation}

A crucial property of such a power law is {scale-invariance}, \emph{i.e.}, when multiplying any of the variables $x_i$ by a constant $a$, the power law in Equation \eqref{eq:power_law} essentially remains the same, being multiplied only by a~constant~$a^{\upbeta_i}$.

In nuclear fusion experiments based on magnetic confinement of a hot hydrogen plasma, scaling laws are crucial for predicting the performance of future fusion reactors, which will have a larger size, magnetic field, plasma density, \emph{etc}., compared to present-day experimental devices~\cite{iter:ch2_07}. These scaling laws can be estimated on the basis of datasets from multiple fusion devices, spanning a significant part of the parameter space. Ordinary least squares regression (OLS) combined with frequentist theory is the statistical workhorse that is employed for this purpose in the vast majority of cases. However, often, there is considerable uncertainty in the experimental data, including the predictor variables, and in the model equations (regression model). However, OLS only deals with uncertainty on the response variables and does not cover additional complications, including atypical observations (outliers), heteroscedasticity, correlations, non-Gaussian distributions, \emph{etc}. As such, OLS regression is often unsuitable for deriving scaling laws~\cite{xiao:log,mcdonald:stat_model}, and many scientific fields could benefit greatly from a unified regression methodology that is flexible and robust and yet relatively simple to implement.

In order to be able to handle the complications mentioned above, we have developed a new regression method, called {geodesic least squares regression} (GLS). It is based on minimization of the Rao geodesic distance between probability distributions on a manifold equipped with the Fisher metric. In this paper, we briefly introduce the method by means of a simple example involving a power law and Gaussian noise. We show the good performance of the method on synthetic data, introducing outliers in the first case and studying the effect of a logarithmic transformation of the data in the second case. Finally, we present an application to the important scaling concerning the power threshold for the transition into the high confinement regime (H-mode) in nuclear fusion experiments based on magnetic plasma confinement. The details of the quantities involved in this scaling, their experimental determination and the underlying physics are not important for the purpose of this paper. Rather, we here aim at showing the performance of GLS on a challenging and heterogeneous real-life dataset.

The paper is structured as follows. The method of geodesic least squares regression is described in Section~\ref{sec:gls}, including a short discussion on calculating geodesic distances on a Gaussian probabilistic manifold, within the framework of information geometry. The next section, Section~\ref{sec:lh}, briefly introduces the database that is used in the subsequent regression experiments, in relation to the scaling law for the H-mode power threshold in fusion plasmas. The experiments involving synthetic data are described in Section~\ref{sec:num}, while the real power threshold scaling is derived in Section~\ref{sec:pow}. Section~\ref{sec:concl} concludes the paper and contains an outlook towards future work related to the methodology.

\section{Geodesic Least Squares Regression} \label{sec:gls}

We start by describing the GLS methodology, which was already introduced in~\cite{verdoolaege:maxent13,verdoolaege:maxent14}, but here, we go into some more detail. We describe a specific form of GLS regression, and it should be stressed that various aspects can be generalized, as will be noted accordingly. Furthermore, several elements on which GLS is based are also found in other regression techniques. The strength of GLS regression is that it integrates many of these aspects in an elegant way, resulting in a method that is very general, flexible and robust. From one point of view, GLS is similar to a class of parameter estimation methods that are collectively referred to in the statistics community as {minimum distance estimation}, in that GLS minimizes a distance between a parametric model distribution of the data and an empirical distribution~\cite{basu:mde}. We use the Rao geodesic distance (GD) as a similarity measure, which has the advantage that it offers an intuitive geometric interpretation. In addition, there are similarities between GLS and the generalized linear model~\cite{mccullagh:glm}.

We will consider the case of regression with multiple predictor variables (regressors) and a single response variable. For this case, we will show that GLS regression can be regarded as a generalization of OLS. However, GLS takes place on a probabilistic manifold, whereas classic OLS operates in a flat Euclidean space. Indeed, OLS is based on minimizing the difference, \emph{i.e.}, the Euclidean distance, between the predicted and measured values of the response variable. Likewise, GLS is based on minimizing the GD between distributions on the probabilistic manifold. Therefore, we start by briefly introducing some concepts from information geometry related to distance calculation.

\subsection{Distance in Information Geometry}

In information geometry, a parametric family of probability densities is interpreted as a Riemannian differentiable manifold~\cite{amari:ig}. Each point on the manifold corresponds to a specific probability density function (pdf) within the family, and the family parameters represent a coordinate system on the manifold. The {Fisher information} (covariance of the score) provides a unique metric tensor. For a probability model $p(\{x_m\}|\{\uptheta^k\}) $ \cite{footnote1} describing a set $\{x_m\}$ of $M$ variables ($m=1,\ldots,M$), parameterized by a set $\{\uptheta^k\}$ of $P$ parameters ($k=1,\ldots,P$), the entries $g_{ij}$ of the Fisher information matrix are given by (no summation):
\begin{equation*}
 \begin{aligned}
  g_{ij}\big(\{\uptheta^k\}\big) &= \mathbb{E}\left[\frac{\p}{\p\uptheta^i}\ln p\big(\{x_m\}|\{\uptheta^k\}\big) \frac{\p}{\p\uptheta^j}\ln p\big(\{x_m\}|\{\uptheta^k\}\big) \right] \\
  &= -\mathbb{E}\left[\frac{\p^2}{\p\uptheta^i \p\uptheta^j}\ln p\big(\{x_m\}|\{\uptheta^k\}\big)\right]
 \end{aligned}, \qquad i,j,k = 1,\ldots, P.
\end{equation*}

The metric provides the basis for distance measurement between pdfs. Specifically, a geodesic curve locally minimizes the distance between two points on the manifold equipped with that metric. Through calculus of variations, it can be shown that a geodesic is the solution of the following system of nonlinear second-order ordinary differential equations, known in the language of variational analysis as {Euler--Lagrange equations}~\cite{oprea:dg} and in the present context as {geodesic equations}:
\begin{equation} \label{eq:geodesic}
 \ddt{\uptheta^r}(t) + \sum_{i,j=1}^P\G{r}{i}{j}\,\dt{\uptheta^i}(t)\dt{\uptheta^j}(t) = 0, \qquad r = 1,\ldots, P.
\end{equation}

Here, the $\uptheta^i$ have been parameterized along the geodesic by $t$ and $\G{r}{i}{j}$ are the \mbox{Christoffel} symbols of the second kind, defined through the metric as:
\begin{equation} \nonumber
 \G{k}{i}{j} = \frac{1}{2} \sum_r g^{kr}\left(\frac{\p g_{jr}}{\p\uptheta^i} + \frac{\p g_{ir}}{\p\uptheta^j}
  - \frac{\p g_{ij}}{\p\uptheta^r}\right),
\end{equation}
where $g^{ij}$ denotes the components of the inverse metric. The boundary value problem Equation \eqref{eq:geodesic} needs to be solved assuming the known values of the coordinates at the boundary points of the geodesic.

From the metric and the solution of the geodesic equations, the length $L_\mathrm{g}$ of the geodesic curve between two distributions with parameter sets $\{\uptheta_1^i\}$ and $\{\uptheta_2^i\}$, \emph{i.e.}, the geodesic distance between these distributions, may be locally calculated as follows (assuming $t$ runs from zero to one):
\begin{equation} \label{eq:geo_dist}
 L_\mathrm{g} = \int_{\{\uptheta_1^i\}}^{\{\uptheta_2^i\}}\ud s = \int_0^1 \left(\sum_{i,j}g_{ij}\dt{\uptheta^i}\dt{\uptheta^j}\right)^{1/2}\; \ud t,
\end{equation}
where $s$ represents the arc length. In the framework of information geometry, the geodesic distance based on the Fisher metric is often referred to as the {Rao geodesic distance} (GD).

Coming back to Equations \eqref{eq:geodesic} and \eqref{eq:geo_dist}, it should be noted that closed-form expressions for the GD are rarely available. On the other hand, provided the Fisher metric can be calculated relatively easily, the framework of information geometry is very useful, since straightforward approximations of the geodesic curves can be found in a geometrically intuitive way~\cite{verdoolaege:jmiv11}. This intuitive approach by means of geometry is an important and attractive aspect of the theory, as it provides enhanced insight into various concepts and algorithms in probability theory and statistics~\cite{kass:ig}. This is also the case for GLS, as will be demonstrated below. Furthermore, as far as the GD is concerned, visualization of geodesics may guide controlled approximations to the geodesic paths and geodesic distances~\cite{verdoolaege:jmiv11}.

Besides the attractive feature of providing intuitive geometrical insight into problems involving similarity measurement between probability distributions, the GD has several more advantages over other similarity measures for distributions. First, it is a distance measure (a metric) in the strict sense of the word. As a result, it is symmetric in its arguments, a desirable property for measuring the similarity between two given states of information in terms of probability distributions. In addition, it obeys the triangle inequality, yielding various practical advantages, for instance in the field of data retrieval from large databases~\cite{verdoolaege:ijcv11}. Furthermore, closed-form expressions may be available for the GD, or its approximation, for various families of distributions where no such analytic form has been found in the case of, for instance, the Kullback--Leibler divergence (KLD)~\cite{verdoolaege:jmiv11}. Finally, there is considerable experimental evidence suggesting that the GD in general is a more effective similarity measure between distributions than the KLD (see~\cite{verdoolaege:jmiv11} and the references therein). We note that for distributions that lie infinitesimally close on the probabilistic manifold, it can be proven that the Kullback--Leibler divergence equals half of the squared geodesic distance between the distributions (see, e.g.,~\cite{kullback:inft}). Hence, in such a case, the KLD and GD yield similar results, but in general, they are quite different measures of similarity between distributions.

\subsection{Geodesics for the Univariate Normal Distribution}

In this paper, we discuss applications that are based on a univariate normal distribution $\mathcal{N}\left(\upmu,\upsigma^2\right)$, parameterized by its mean $\upmu$ and standard deviation $\upsigma$. In this case, an analytic expression for the Fisher--Rao metric is available. It turns out to be the familiar Poincar\'e metric, which, when applied to a half-plane, is a well-known model for hyperbolic geometry that has constant negative scalar curvature. The line-element is given by~\cite{atkinson:rao,burbea_rao:diff_metric}:
\begin{equation} \label{eq:line_element}
 \ud s^2 = \frac{\ud\upmu^2}{\upsigma^2} + 2\frac{\ud\upsigma^2}{\upsigma^2}.
\end{equation}

\begin{figure}[H]
	\centering
 \includegraphics[width=0.45\textwidth]{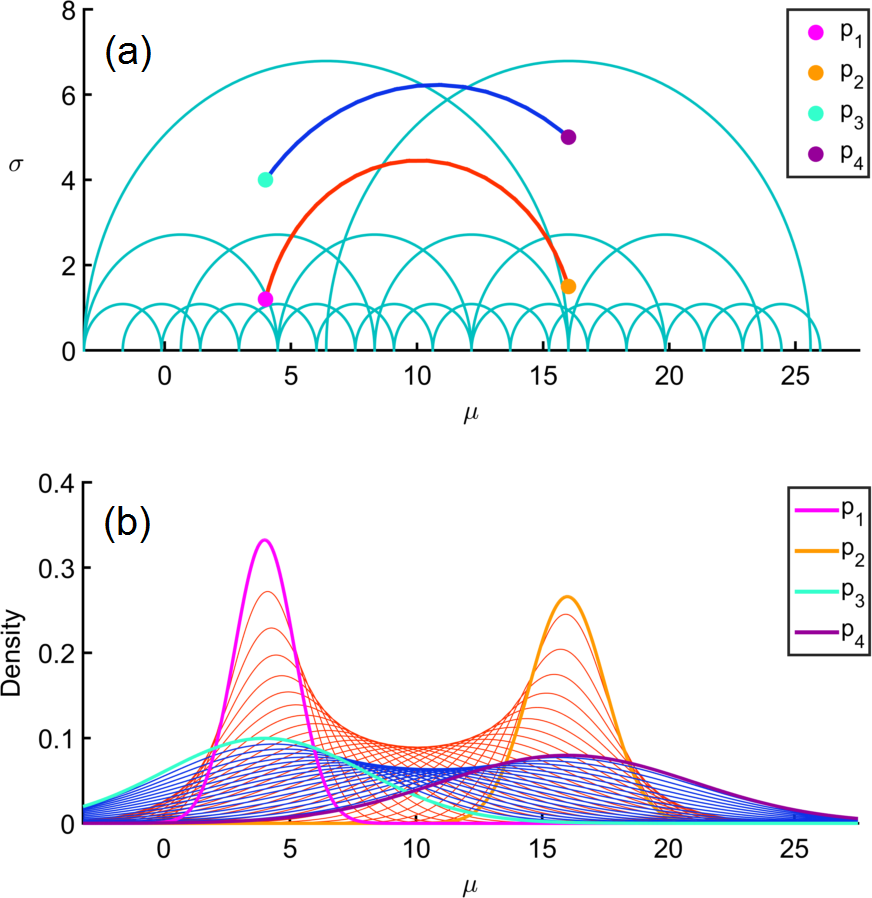} \\[0.5cm] \includegraphics[width=0.25\textwidth]{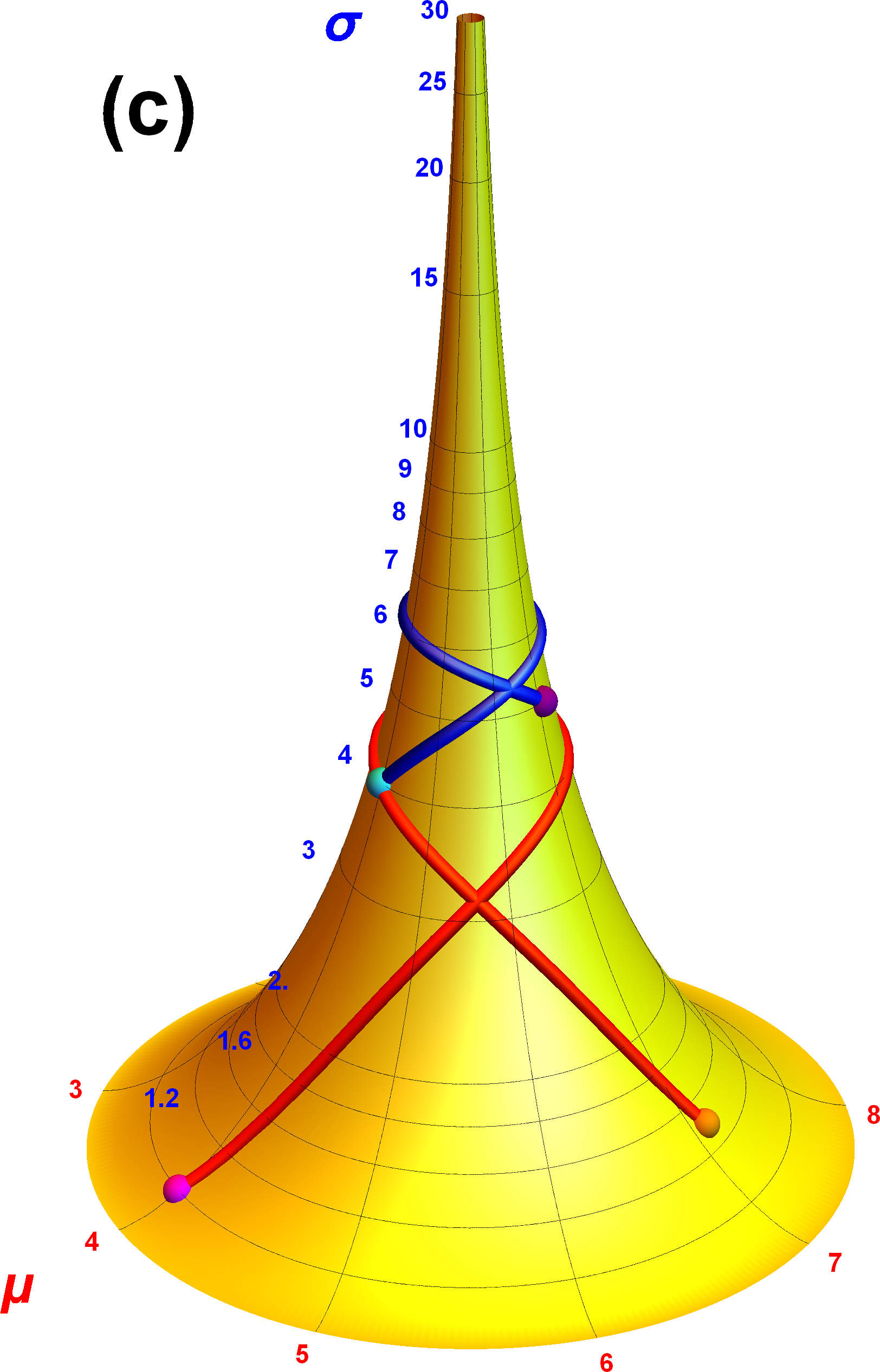} \qquad
 \includegraphics[width=0.24\textwidth]{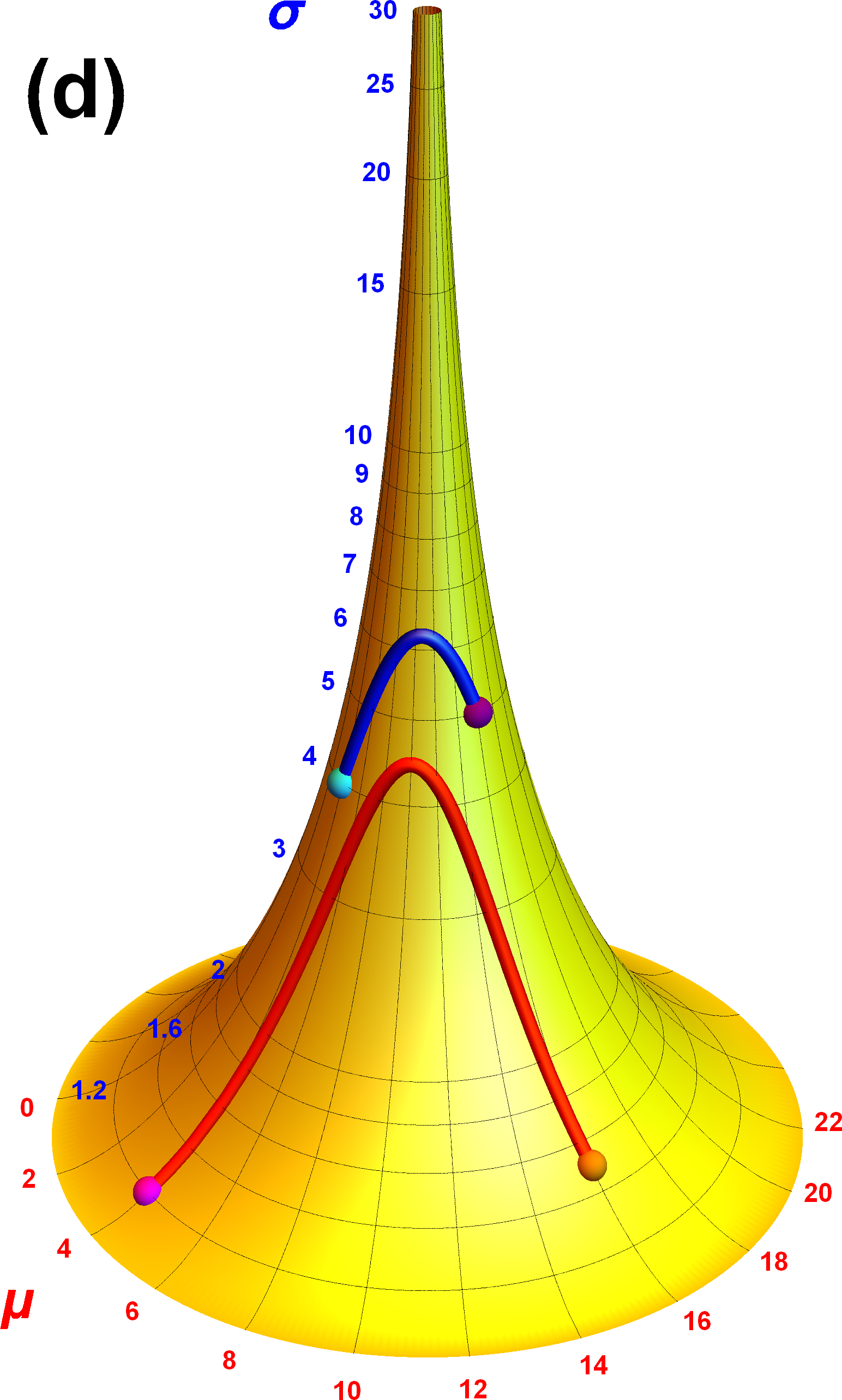}
 \caption{(\textbf{a}) Illustration of the Poincar\'e half-plane with several half-circle geodesics in the background, together with the geodesic between the points $p_1$ and $p_2$ and between $p_3$ and $p_4$, defined in the main text. (\textbf{b}) Probability densities corresponding to the points $p_1$, $p_2$, $p_3$ and $p_4$ indicated in ({a}). The densities associated with some intermediate points on the geodesics between $p_1$ and $p_2$ and between $p_3$ and $p_4$ are also drawn. (\textbf{c}) Rendering of one blade of the tractroid, again with the two geodesics superimposed. The parallels of the tractroid are lines of constant standard deviation $\upsigma$, while the meridians (the tractrices) are lines of constant mean $\upmu$. This representation of the normal manifold is periodic in the $\upmu$-direction, and a rescaled version (longer period along $\upmu$) is shown in (\textbf{d}).\label{fig:hyperbolic}}
\end{figure}

As an illustration, the Poincar\'e half-plane is pictured in Figure~\ref{fig:hyperbolic}a, together with two geodesics between, on the one hand, the points $p_1 = \mathcal{N}\left(4,1.2^2\right)$ and $p_2 = \mathcal{N}\left(16,1.5^2\right)$ and, on the other hand, $p_3 = \mathcal{N}\left(4,4.0^2\right)$ and $p_4 = \mathcal{N}\left(16,5.0^2\right)$. The corresponding normal density functions are drawn in Figure~\ref{fig:hyperbolic}b, as well as a number of densities associated with some intermediate points on each geodesic. As a further illustration, Figure~\ref{fig:hyperbolic}c shows one blade of a particular {pseudosphere}, namely the tractroid, which is locally isometric to the Poincar\'e half-plane and the univariate normal manifold for $\upsigma>1$, with periodicity in $\upmu$. In order to better visualize the geodesics, a rescaled version of the tractroid is shown in Figure~\ref{fig:hyperbolic}d. This surface has a longer period in the $\upmu$-direction. However, it should be kept in mind that only the visualization in Figure~\ref{fig:hyperbolic}c can be used to measure absolute distances on the surface, for in Figure~\ref{fig:hyperbolic}d, the pictured geodesics are no longer the shortest curves between the points in question. \mbox{It is clear} that the geodesics on the Gaussian manifold are different from straight lines in the Euclidean space, wherein the manifold has been immersed. The shape of the geodesics can be made intuitively clear by noting that they always pass through a region of increased standard deviation relative to that of the boundary points. This provides the shortest route, as can be seen from the line element Equation~\eqref{eq:line_element}. Interestingly, similar arguments will be shown to enable a deeper insight into the operation of GLS regression. We further note that various alternative models exist to visualize hyperbolic geometry; see,~e.g.,~\cite{nielsen:socg14}.

A closed-form expression is available for the GD on the normal manifold, permitting fast evaluation. Indeed, for two univariate normal distributions $p_1\left(x|\upmu_1,\upsigma_1^2\right)$ and $p_2\left(x|\upmu_2,\upsigma_2^2\right)$, the GD is given by~\cite{burbea_rao:diff_metric}:
\begin{equation} \label{eq:gd}
 \mathrm{GD}(p_1,p_2) = \sqrt{2}\ln\frac{1+\updelta}{1-\updelta} = 2\sqrt{2}\tanh^{-1}\updelta,
 \qquad \updelta\equiv\left[\frac{(\upmu_1-\upmu_2)^2 + 2(\upsigma_1-\upsigma_2)^2}{(\upmu_1-\upmu_2)^2 + 2(\upsigma_1+\upsigma_2)^2}\right]^{1/2}.
\end{equation}

Furthermore, since the injectivity radius of the hyperbolic plane is infinite, the geodesics are globally length-minimizing~\cite{oprea:dg}.

\subsection{Geodesic Least Squares Methodology}

GLS starts from the premise that the probability distribution underlying experimental measurements is the fundamental object resulting from the measurement. As such, GLS does not perform regression based on data points in a Euclidean space, but rather operates on probability distributions lying on a probabilistic manifold. This introduces additional flexibility that renders the method robust in the presence of large uncertainties, as will be demonstrated in the experiments.

Briefly, the idea is to consider two different proposals for the distribution of the response (dependent) variable $y$, conditional on the predictor variables. On the one hand, there is the distribution that one would expect if all assumptions were correct regarding the deterministic component of the regression model (regression function) and the stochastic component. We call this the {modeled distribution}. On the other hand, we try to capture the conditional distribution of $y$ by relying less on the model assumptions, but directly on the measurements of $y$. For this, we will use the term {observed distribution}. It is in this sense that GLS is similar to minimum distance estimation (MDE), where the Hellinger distance is a popular similarity measure~\cite{beran:mde}. This was first applied to regression in~\cite{pak:mde_reg}, but there are several differences with GLS. First and foremost, GLS calculates the geodesic distance between each {individual} pair of modeled and observed distributions of the response variable, corresponding to an individual measurement point. As such, each individual data point acquires the status of a probability distribution in its own right. Consequently, GLS performs regression between probability distributions on a probabilistic manifold. In contrast, MDE usually considers a distance between a kernel density estimate of the distribution of residuals, on the one hand, and the parametric model, on the other hand, based on the entire data sample. Secondly, we explicitly model all parameters of the modeled distribution, which is similar to the ideas behind the link function in the generalized linear model (GLM)~\cite{mccullagh:glm}. In the present work, this will be accomplished by explicitly modeling both the mean and standard deviation of the Gaussian modeled distribution. Additionally, a final difference is that we use the Rao geodesic distance as a similarity measure.

As a simple example that we will use also in the experiments, consider a linear relation $\upeta = \upbeta\xi$ between a single predictor variable $\upxi$ and a response variable $\upeta$, with $\upbeta$ a constant. In accordance with the discussion above, we explicitly wish to allow for the challenging case of uncertainty on the predictor variable $\upxi$. Therefore, we assume that, in reality, $N$ samples of a stochastic (noisy) variable $x$ are observed, together with $N$ samples of a stochastic response variable $y$. We take the simple case of normally distributed (Gaussian) noise:
\begin{equation} \label{eq:noise}
 \begin{aligned}
  y &= \upeta + \epsilon_y = \upbeta\upxi + \epsilon_y, &\qquad\epsilon_y \sim \mathcal{N}\left(0, \upsigma_y^2\right), \\
  x &= \upxi + \epsilon_x, &\qquad\epsilon_x \sim \mathcal{N}\left(0, \upsigma_x^2\right).
 \end{aligned}
\end{equation}

The observations $x_n$ ($n=1,\ldots,N$) are taken as mutually independent and so are the $y_n$. $\upsigma_x$ and $\upsigma_y$ are assumed to be known, and in this example, they are taken constant for all measurements, \emph{i.e.}, we have homoscedasticity. However, we will also consider heteroscedasticity later on. According to the regression model, conditionally on $x_n$, each measurement $y_n$ is drawn from a normal distribution:
\begin{equation} \label{eq:cond_distr}
 p_\mathrm{mod}(y|x_n) = \mathcal{N}\left(\upbeta x_n,\upsigma_\mathrm{mod}^2\right), \quad \text{where} \quad \upsigma_\mathrm{mod}^2 \equiv \upsigma_y^2 + \upbeta^2\upsigma_x^2,
\end{equation}
with the subscript ``mod'' referring to the modeled distribution. In our simple example, Equation~\eqref{eq:cond_distr} follows from standard Gaussian error propagation rules. However, for nonlinear regression laws, the conditional distribution for $y$ has to be obtained by marginalizing the unknown true values $\upxi_n$. Nevertheless, the Gaussian error propagation laws may be used in the nonlinear case as well, to {approximate} the conditional distribution $p(y|x_n)$ by a normal distribution, as will be shown \mbox{in the experiments.}

We next choose a specific form of the observed distribution corresponding to each realization of the variable $y$, conditional on the observations, \emph{i.e.}, $p_\mathrm{obs}(y|y_n)$. In this example, we take again the normal distribution, but centered on each data point: $\mathcal{N}\left(y_n,\upsigma_\mathrm{obs}^2\right)$, where $\upsigma_\mathrm{obs}$ is to be estimated from the data. In the context of the GLM, this is known as the {saturated model}. The extra parameter $\upsigma_\mathrm{obs}$ gives the method added flexibility, since it is not \emph{a priori} required to equal $\upsigma_\mathrm{mod}$. As a result, GLS is less sensitive to incorrect model assumptions. Note that in this example, we have chosen the observed distribution from the same model (Gaussian) as the modeled distribution. Furthermore, $\upsigma_\mathrm{mod}$ is taken as a fixed value for all measurements and so is $\upsigma_\mathrm{obs}$. These assumptions can of course be relaxed, leading to a more general method. However, the transition from OLS to GLS is best explained by means of a Gaussian observed distribution, which, in addition, offers computational advantages, since the expression for the GD has a closed form; see~Equation \eqref{eq:gd}.

GLS now proceeds by minimizing the total GD between, on the one hand, the joint observed distribution of the $N$ realizations of the variable $y$ and, on the other hand, the joint modeled distribution. Thanks to the independence assumption in this example, we can write this in terms of products of the corresponding marginal distributions:
\begin{align}
 \hat{\upbeta} &= \underset{\upbeta \in \mathbb{R}}{\mathrm{argmin}} \, \mathrm{GD}\left[\prod_{n=1}^N p_\mathrm{obs}(y|y_n),\prod_{n=1}^N p_\mathrm{mod}(y|x_n)\right] \nonumber \\
 &= \underset{\upbeta \in \mathbb{R}}{\mathrm{argmin}} \, \sum_{n=1}^N\mathrm{GD}^2 \big[p_\mathrm{obs}(y|y_n),p_\mathrm{mod}(y|x_n)\big]. \label{eq:obj_fun}
\end{align}

The last equality entails a considerable simplification, owing to the property that the squared GD between products of distributions can be written as the sum of squared GDs between the corresponding factors~\cite{burbea_rao:diff_metric}. Hence, the optimization procedure involves matching not only $y_n$ with $bx_n$, but also $\upsigma_\mathrm{obs}^2$ with $\upsigma_y^2 + \upbeta^2\upsigma_x^2$. Note that the parameter $\upbeta$ occurs both in the mean and the variance of the modeled distribution. Incidentally, forcing $\upsigma_\mathrm{obs}^2\equiv \upsigma_y^2 + \upbeta^2\upsigma_x^2$ would take us back to standard maximum likelihood estimation, for the Rao GD between the two Gaussians $p_\mathrm{obs}$ and $p_\mathrm{mod}$ with means $y_n$ and $bx_n$, respectively, but with identical standard deviations (fixed along the geodesic path), is precisely the Mahalanobis distance~\cite{rao:diff_metrics}:
\begin{equation*}
 \mathrm{GD}(p_\mathrm{obs},p_\mathrm{mod}) = \frac{|y_n - bx_n|}{\upsigma_y^2 + \upbeta^2\upsigma_x^2}, \qquad \text{if $\upsigma_\mathrm{obs}^2\equiv \upsigma_y^2 + \upbeta^2\upsigma_x^2$}.
\end{equation*}

We note that the GLS scheme addresses many of the difficulties with classic OLS regression. First, GLS explicitly allows uncertainty on the predictor variables, and it is not restricted to normal or symmetric noise distributions, nor does it necessarily assume homoscedasticity. In addition, correlations among variables and among observations can be built into the stochastic component of the regression model. Furthermore, GLS can operate with any (nonlinear) regression function. Moreover, it will be shown in the experiments that GLS is relatively insensitive to uncertainties in both the stochastic and deterministic components of the regression model. The same quality renders the method also robust against outliers.

In the experiments below, we employed a classic active-set algorithm to carry out the optimization~\cite{gill:as}. Furthermore, presently, the GLS method does not directly offer confidence (or credible) intervals on the estimated quantities. Future work will address this issue in more detail, but for now, error estimates were derived by Monte Carlo sampling in the case of the numerical simulations (Section~\ref{sec:num}) and by bootstrapping in the case of the real data (Section~\ref{sec:pow})~\cite{casella:stat}. The bootstrapping involved creating, from the measured dataset, a large number of artificial datasets of the same size, by resampling with replacement. The regression analysis was then carried out on each of the datasets, and the mean and standard deviation, over all datasets, of each estimated regression parameter and of the predicted quantities were used as estimates of the parameter or prediction value and its error bar, respectively. This~scheme typically results in rather conservative error bars, which could possibly be narrowed down using more sophisticated methods.

\section{The L-H Power Threshold and Database} \label{sec:lh}

We now provide some background information regarding the main regression application that will be treated in the experiments with synthetic and real data. It concerns one of the most important scaling relations in fusion science based on magnetic plasma confinement, related to the threshold $P_\mathrm{thr}$ for the heating power that is required for the plasma to make the transition into a desired regime of high energy confinement (H-mode) in the next-step fusion device ITER (International Thermonuclear Experimental Reactor)~\cite{iter:ch2_07,snipes:iaea02,martin:power08}. To a good approximation, this so-called {L-H} (or {H-mode}) {power threshold} depends on the electron density in the plasma $\bar{n}_\mathrm{e}$ (in $10^{20}$~$\mathrm{m}^{-3}$), the main magnetic field $B_\mathrm{t}$ (in tesla (T)) and the total surface area $S$ of the confined plasma (in $\mathrm{m}^2$). This~is usually expressed by means of the following scaling relation:
\begin{equation} \label{eq:power}
 P_\mathrm{thr} = \upbeta_0\, \bar{n}_\mathrm{e}^{\upbeta_1} B_\mathrm{t}^{\upbeta_2} S^{\upbeta_3}.
\end{equation}

To estimate the coefficients in this relation, we employed data from eight fusion devices of the tokamak type (ASDEX
, AUG
, CMOD
, DIII-D
, JET
, JFT-2M
, JT-60U
, PBXM
) in the International Tokamak Physics Activity (ITPA) multi-machine database for the L-H power threshold (subset IAEA02
~\cite{ryter:lh_database,ryter:lh_database_progress,snipes:iaea02,itpa:power_iaea02}). This yields $616$ measurement sets of power, density, magnetic field and surface area, each set obtained during a brief time of plasma operation under stationary conditions in one of the eight devices involved in the study \cite{footnote2}.

The ITPA database contains some information regarding the error bars on the measurements, specifically relative errors expressed as percentages. This is important for our purposes, because we need the error bars to calculate $\upsigma_\up{mod}$. Unfortunately, the error estimates are not available in some cases, and if they are, the precise definition of the error bars is not always clear. Usually, an error bar in the database represents an estimate by the experimentalist of the typical range in which the ``true'' quantity can be expected to lie, where the uncertainty is assumed to be caused by {both} stochastic and systematic effects. Moreover, it is difficult to assess the probability that is covered by the stochastic component of the errors mentioned in the database. Since a detailed investigation of the uncertainty of the threshold data is beyond the scope of the present paper, we will assume that the error bars pertain to a stochastic uncertainty corresponding to a single standard deviation of a Gaussian distribution. For some derived quantities, the error bars had to be calculated from the uncertainty on more fundamental measurements. In those cases, we employed Gaussian error propagation rules to estimate the standard deviation on the derived quantities. For the case of the global H-mode confinement database, this strategy has been shown to provide reasonable information on the actual error bars~\cite{verdoolaege:ppcf12}.

It is important to mention that the main source of uncertainty in the data used for power threshold scaling, when compared to the predictions of a simple power law regression model, is not expected to be due to the measurement uncertainty on the individual variables. There are far larger sources owing to the variability of the experiments that produced the data. To estimate the variability of each of the physical quantities with respect to the scaling law, we performed the following calculation. First, the nonlinear scaling law was estimated using OLS, as explained in Section~\ref{sec:pow_nlin}. Then, for a specific variable $z$ (one of the predictor variables or the dependent variable) and for each data point, the {relative} difference was computed between the $z$-value of the data point itself and the $z$-value of the projection of the data point on the hypersurface given by the scaling law, keeping the values of the other variables fixed. This~difference can be interpreted as the deviation of the point from the theoretical scaling law, assuming the deviation is solely due to the variability of the variable $z$. Finally, the standard deviation of these relative differences was taken, and the procedure was repeated for every predictor variable and the dependent variable. The resulting standard deviations can be interpreted as upper bounds of the relative variability of each of the variables around their `theoretical' values given by the scaling law. This way, for $\bar{n}_\mathrm{e}$, $B_\mathrm{t}$, $S$ and $P_\mathrm{thr}$, we obtained $39$\%, $31$\%, $28\%$ and $38\%$, respectively. These levels are clearly much larger than the relative uncertainties due to measurement error alone. Indeed, the typical measurement error bars quoted in the ITPA database, on average, over all devices, are estimated at $4$\% for $\bar{n}_\mathrm{e}$, $1$\% for $B_\mathrm{t}$, $3$\% for $S$ and $15\%$ for $P_\mathrm{thr}$~\cite{ryter:lh_database,ryter:lh_database_progress}.

\section{Numerical Simulations} \label{sec:num}

We next demonstrate some of the potential of the GLS regression scheme by means of a number of experiments with synthetically-generated data. We treat two particular cases of deviation from the model according to which the data were created and show that, in comparison with a number of standard regression techniques, GLS yields the most accurate results across all experiments. The first case concerns the effect of outliers, while in the second case, the influence of a logarithmic transformation is studied. In each case, we start with a very simple experiment that is easily reproduced, using a single predictor variable, providing some intuitive feeling regarding the performance of the method. We then proceed to a more elaborate test, still based on partly synthetic data, but using a regression challenge similar to that used in the real-world experiment for scaling of the L-H power threshold in fusion plasmas, presented in Section~\ref{sec:pow}.

\subsection{Effect of Outliers}

The robustness of minimum distance estimators to outliers in the data was noted in the classic literature of minimum distance estimation~\cite{beran:mde}. We now show that this is a quality also enjoyed by GLS regression.

\subsubsection{Single Predictor Variable}

We first concentrate on estimating the slope of a regression line with a single predictor variable. To this end, a dataset was generated consisting of ten points labeled by coordinates $\upxi_n$ and \mbox{$\upeta_n$ ($n=1,\ldots,10$)}, with the $\upxi_n$ chosen unevenly between zero and 50 and $\upeta_n = \upbeta\upxi_n$, taking $\upbeta=3$. Then, Gaussian noise was added to all coordinates according to Equation \eqref{eq:noise}, with $\upsigma_y = 2.0$ and $\upsigma_x = 0.5$, resulting in values $x_n$ for the predictor variable and $y_n$ for the response variable. Finally, one outlier was created by multiplying the value of $y_k$ by a factor distributed uniformly between $1.5$ and $2.5$, with $k$ chosen uniformly among the indices 8, 9 and 10.

We next estimated $\upbeta$ by means of GLS and compared the estimates with those obtained by OLS, maximum \emph{a posteriori} (MAP) using the model in Equation~\eqref{eq:cond_distr} for the likelihood and an uninformative prior~\cite{preuss:all_variables}, total least squares (TLS), which is a typical errors-in-variables technique~\cite{markovsky:tls}, and a robust method (ROB) based on iteratively reweighted least squares (bisquare weighting)~\cite{maronna:robust}, included in the MATLAB Statistics toolbox~\cite{mathworks:matlab_stat}. It should be noted that MAP takes into account the error bars on the predictor variables. In all cases, we assumed knowledge of the values of $\upsigma_x$ and $\upsigma_y$. In order to get an idea of the variability of the estimates, Monte Carlo sampling of the data-generating distributions was performed, and the estimation was carried out $100$ times.

The results are given in Table~\ref{tab:reg_outlier_stat}, mentioning the sample average and standard deviation of the estimates $\hat{\upbeta}$ over the $100$ runs for each of the methods. GLS is seen to perform very well and similar to the robust method ROB, but the other techniques yield considerably worse results. The average estimate of $\upsigma_\mathrm{obs}$ was $5.43$ with a standard deviation of $0.24$. On the other hand, the modeled value of the standard deviation in the conditional distribution for $y$ was $\upsigma_\up{mod}=\sqrt{\upsigma_y^2 + 9\upsigma_x^2}=2.5$. Hence, GLS succeeds in ignoring the outlier by increasing the estimated variability of the data. Put differently, the effect of the outlier is, in a sense, to increase the overall variability of the data, which GLS takes into account by increasing the observed standard deviation of the data ($\upsigma_\mathrm{obs}$) with respect to the standard deviation predicted by the model ($\upsigma_\mathrm{mod}$).

\begin{table}[H]
 \caption{Monte Carlo estimates of the mean and standard deviation for the slope parameter in linear regression with errors on both variables and one outlier. GLS, {geodesic least squares regression}; TLS, total least squares; ROB, robust method.\label{tab:reg_outlier_stat}}
 \centering
 \small
 \begin{tabular}{cccccc}	
 	\toprule
 	 \textbf{Original} & \textbf{GLS} & \textbf{OLS}   & \textbf{MAP}   & \textbf{TLS}  & \textbf{ROB} \\
  	\midrule
 	 $\upbeta=3.00$ & \textbf{3.031} $\bm{\pm}$ \textbf{0.035} & 3.68 $\pm$ 0.29 & 3.83 $\pm$ 0.36 & 4.6 $\pm$ 1.0 & 2.992 $\pm$ 0.041 \\
 	\bottomrule
 \end{tabular}
\end{table}

\begin{figure}[H]
	\centering
 \includegraphics[width=0.5\textwidth]{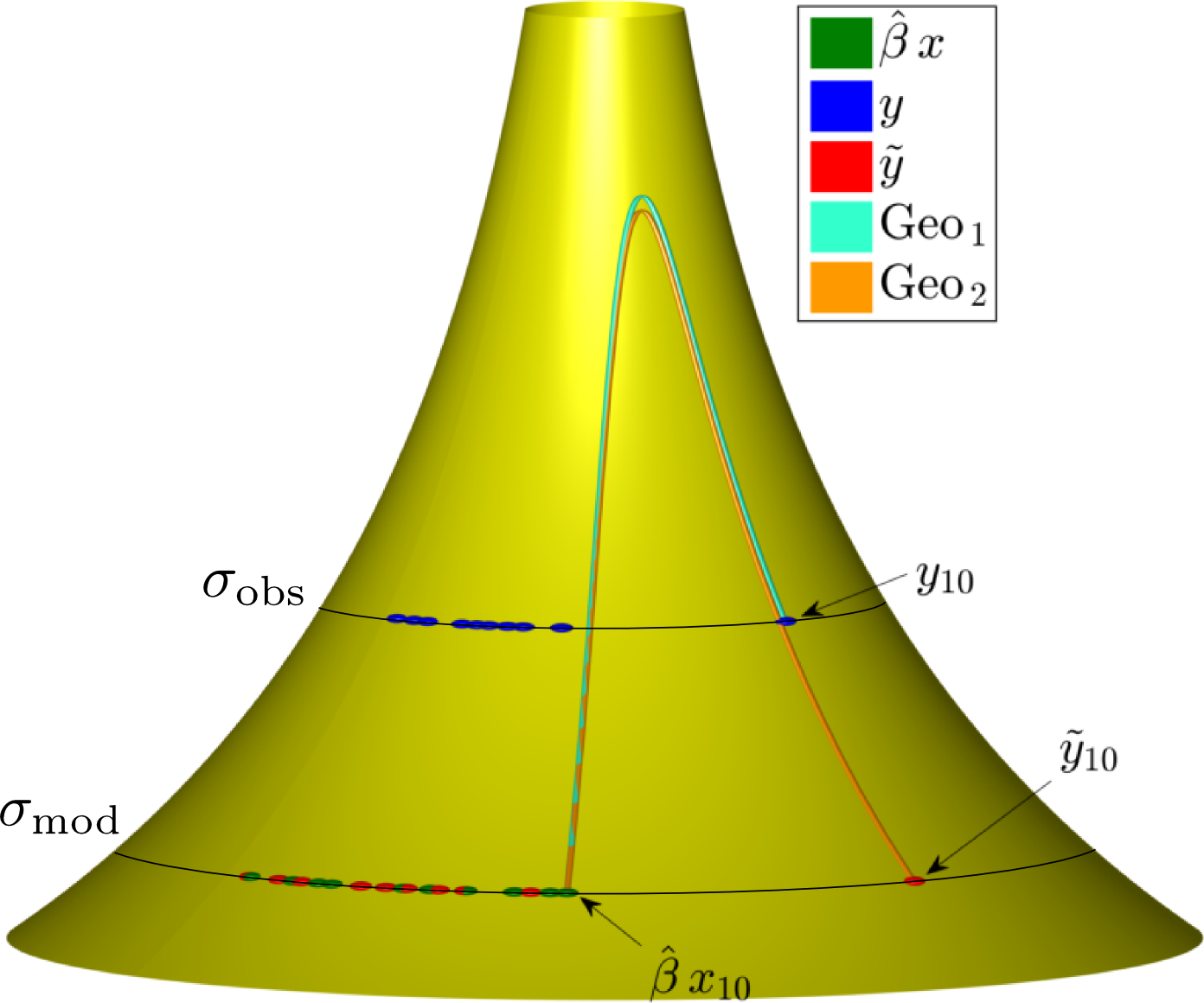}
 	\caption{A portion of the pseudosphere together with the regression results on synthetic data with an outlier, as described in the main text.\label{fig:ps_data}}
\end{figure}

As mentioned before, this result can also be understood in terms of the pseudosphere as a geometrical model for the normal distribution. To see this, we refer to Figure~\ref{fig:ps_data}, where several sets of points (distributions) are drawn on a portion of the surface of the pseudosphere for one particular dataset generated as described above. First, the modeled distributions are plotted with their means $\hat{\upbeta}x_n$ (see~Equation \eqref{eq:cond_distr}) and standard deviations $\upsigma_\mathrm{mod}=2.5$, using the average estimate $\hat{\upbeta}=3.031$ obtained by GLS. These are the green points on the surface, and they lie on a parallel, since they all correspond to Gaussians with the same standard deviation $\upsigma_\mathrm{mod}$. In this particular dataset, the index of the outlier was $k=10$, so the point $\hat{\upbeta}x_{10}$ is indicated individually. Obviously, according to the model, no outlier is expected, so the modeled distribution corresponding to $k=10$, which is the green point just mentioned, lies close to the other predicted points (distributions). Next, we plot the observed distributions with their means $y_n$ and standard deviations $\upsigma_\mathrm{obs}$ (for this dataset estimated at $\hat{\upsigma}_\mathrm{obs}=5.43$). These are the blue points, lying at a constant standard deviation $\upsigma_\mathrm{obs}$, which is higher than $\upsigma_\mathrm{mod}$ ($5.43 > 2.5$). The outlier $y_{10}$ can clearly be observed, and being an outlier, it lies relatively far away from the rest of the blue points (observed distributions). Now suppose that, like MAP, GLS would {not} be able to increase $\upsigma_\mathrm{obs}$ relative to $\upsigma_\mathrm{mod}$ in order to accommodate the outlier. Then, the observed distributions would have the same observed means (the measured values $y_n$), but they would have the standard deviation predicted by the model. Hence, they would lie on the parallel corresponding to $\upsigma_\mathrm{mod}$, just like the green points. We have plotted these fictitious distributions as the red points at the level of $\upsigma_\mathrm{mod}$, and they are labeled $\tilde{y}$. Again, the outlier (labeled $\tilde{y}_{10}$) can be seen, but it seems to lie {further} away from the other red points (the points $\tilde{y}_n$) compared to the actually observed situation, \emph{i.e.}, the distance from $y_{10}$ to the other $y_n$ (blue points). At least this is the case when using (visually) the Euclidean distance in the embedding Euclidean space. We can verify that this is indeed so by using the proper geodesic distance on the surface: overall, the blue points lie closer together (including the outlier) than the red points. Now, in fact, GLS aims at minimizing the distance between each green point (modeled distribution) and its corresponding blue point (observed distribution), so as far as the outlier is concerned, we should really be looking at the geodesic between the point $(\hat{\upbeta}x_{10},\upsigma_\mathrm{mod})$ and the point $(y_{10},\upsigma_\mathrm{obs})$. The geodesic (labeled ``$\mathrm{Geo}_1$'') between these points is also drawn on the surface, and again, we compare this to the fictitious situation, represented by the geodesic (labeled ``$\mathrm{Geo}_2$'') between $(\hat{\upbeta}x_{10},\upsigma_\mathrm{mod})$ and $(\tilde{y}_{10},\upsigma_\mathrm{mod})$. Indeed, again, we see that the geodesic $\mathrm{Geo}_1$ is shorter than $\mathrm{Geo}_2$. Therefore, by increasing $\upsigma_\mathrm{obs}$ relative to $\upsigma_\mathrm{mod}$, the outlier is not so much an outlier anymore, as measured on the pseudosphere!
 When calculating the GD, one finds $2.4$ for $\mathrm{Geo}_1$ and $2.8$ for $\mathrm{Geo}_2$. Therefore, GLS obtains a lower value of the objective function (sum of squared geodesic distances) if it increases $\upsigma_\mathrm{obs}$ with respect to $\upsigma_\mathrm{mod}$. Of course, there is a limit to this: GLS cannot continue raising $\upsigma_\mathrm{obs}$ indefinitely, trying to mitigate the distorting effect of the outlier, for then, the other points would get a too high observed standard deviation, which is not supported by the data. The image that we see in Figure~\ref{fig:ps_data} is the best compromise that GLS could find. In fact, we note that, in the case we suspect that $y_{10}$ could be an outlier, it may very well be worthwhile to introduce {two} parameters to describe the observed standard deviation: one for the nine points that seem to follow the model and one to take care of the outlier. This would be a very straightforward extension of the method, and we explore this to some extent when using data from the ITPA database below. There, we assign a separate parameter to describe the observed standard deviation of all data coming from a specific tokamak, hence defining an individual parameter for each machine.

\subsubsection{Multiple Predictor Variables} \label{sec:outlier_mult}

In this experiment, a regression problem with multiple predictor variables and a power law is studied. The deterministic part of the regression model is based on the real-world problem for the L-H power threshold in fusion plasmas, which we are going to consider in Section~\ref{sec:pow}. Furthermore, the values of the predictor variables are taken from the same international power threshold database, and values of the response variable are synthetically generated from this.

More specifically, the dataset for this experiment was created as follows. First, an artificial linear regression law was put forward for a variable $\upeta$, depending on the predictor variables $\bar{n}_\mathrm{e}$, $B_\mathrm{t}$ and $S$, which were introduced in the context of the power threshold scaling law in Section~\ref{sec:lh} \cite{footnote3}. In particular, we generated a number of realizations of the variable $\upeta$ from the following prescription:
\begin{equation} \label{eq:eta_lin}
 \upeta = \upbeta_0 + \upbeta_1\bar{n}_\mathrm{e} + \upbeta_2 B_\mathrm{t} + \upbeta_3 S.
\end{equation}

This was considered as the ``true'' relation between the predictor and response variables, where, as mentioned above, the values of the predictor variables were chosen to be exactly those from the ITPA database, which are normally used in the real power threshold scaling law. A whole range of datasets was created using the following values of the coefficients $\upbeta_0$, $\upbeta_1$, $\upbeta_2$ and $\upbeta_3$:
\begin{equation} \label{eq:range_beta}
 \begin{aligned}
  \upbeta_0 &= 1,1.1,\ldots,20, \\
  \upbeta_1,\upbeta_2,\upbeta_3 &= 0.1,0.2,\ldots,2.
 \end{aligned}
\end{equation}

Thus, for each combination of values of $\upbeta_0$, $\upbeta_1$, $\upbeta_2$ and $\upbeta_3$, all $616$ values of $\upeta$ were calculated according to Equation \eqref{eq:eta_lin}, based on the values of $\bar{n}_\mathrm{e}$, $B_\mathrm{t}$ and $S$ from the ITPA database. The range of coefficient values in Equation \eqref{eq:range_beta} was chosen to be representative for the values that are typically obtained from a regression analysis on the true scaling law (see Section~\ref{sec:pow}). The exception is $\upbeta_0$, for which the range was chosen of roughly the same order as $\upeta - \upbeta_0$ (much smaller values of $\upbeta_0$ would not be estimable in comparison with $\upeta - \upbeta_0$).

Next, Gaussian noise was added to both the predictor and response variables. The noise level was chosen according to the typical relative measurement errors in the ITPA database, \emph{i.e.}, $4$\% for $\bar{n}_\mathrm{e}$, resulting in a variable $x_1$, $1$\% for $B_\mathrm{t}$ (variable $x_2$), $3$\% for $S$ (variable $x_3$) and $15$\% for the dependent variable (variable $y$, which is $P_\mathrm{thr}$ in the real-world regression problem). It should be stressed that, in the light of our comments in Section~\ref{sec:lh} regarding the variability of the predictor variables, these are rather low noise levels. We further note that fixed relative noise levels lead to a different standard deviation for each measurement (heteroscedasticity).

Furthermore, in this experiment studying the effect of atypical observations, $10$ outliers were created in each of the datasets. In particular, from the total of $616$ points in each dataset, $10$ points were randomly chosen, and the associated value of $y$ was multiplied with a factor $F$, where $F$ was distributed uniformly between $1.5$ and $2.5$. For each combination of coefficient values $\upbeta_i$ ($i=0,\ldots,3$) taken from Equation \eqref{eq:range_beta}, $10$ datasets were realized, each time performing the sampling of noise and outliers.

\begin{figure}[H]
	\centering
 \includegraphics[width=0.54\textwidth]{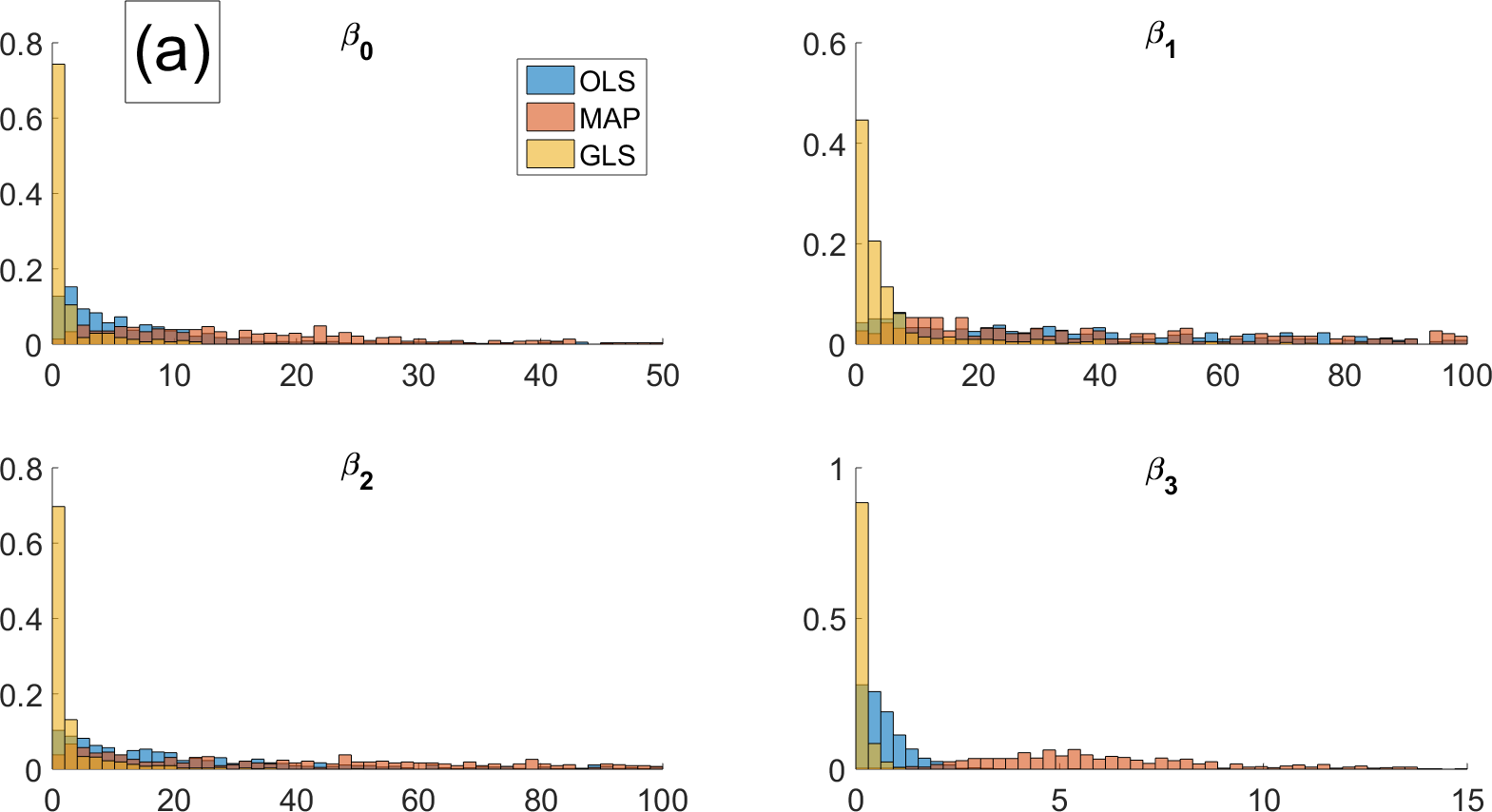}\\
 \includegraphics[width=0.54\textwidth]{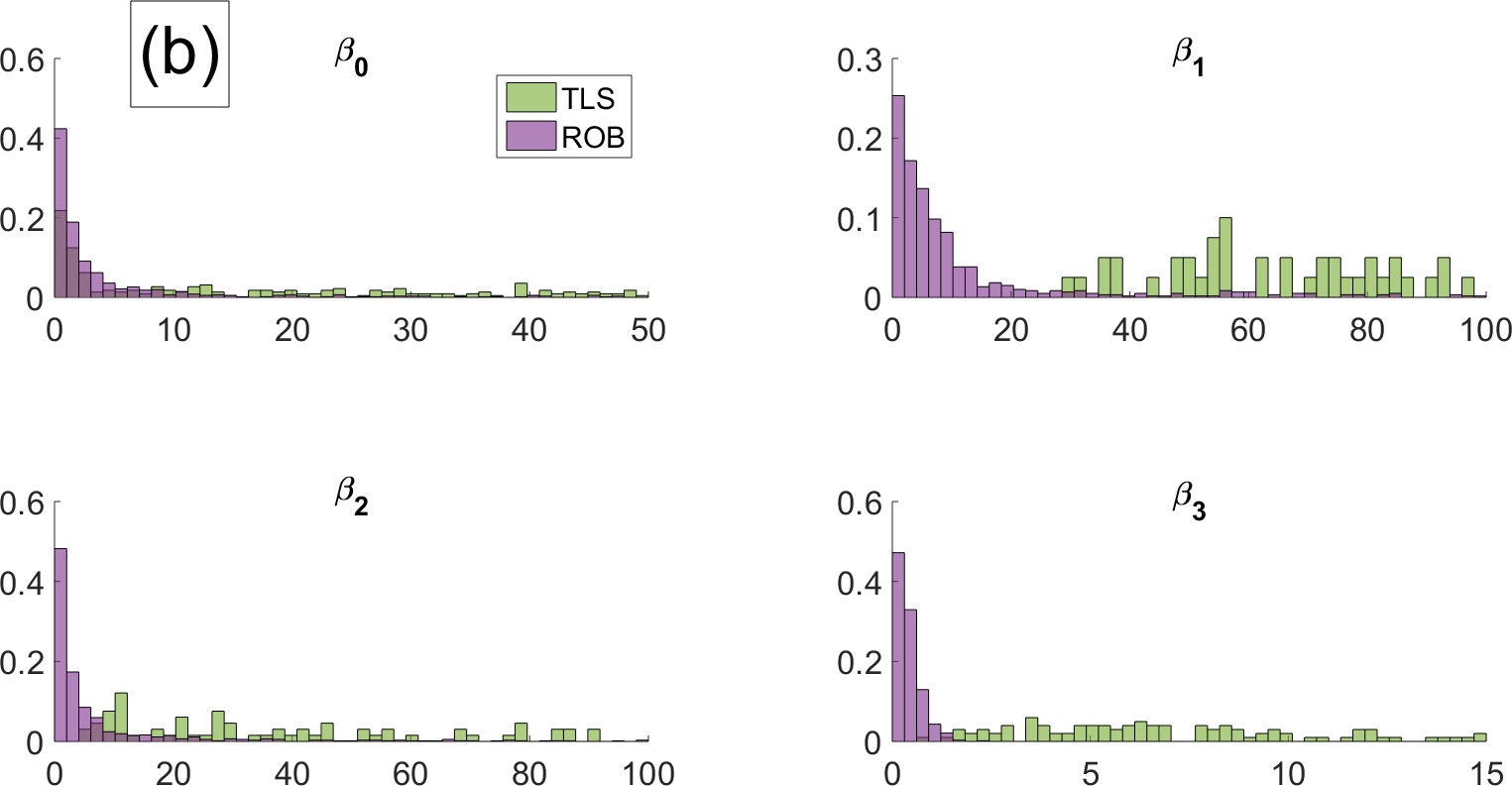}\vspace{12pt}
 \caption{(\textbf{a}) Histograms of the relative error in estimating the regression coefficients $\upbeta_i$ by means of OLS, MAP and GLS for a linear regression problem with outliers. Horizontal axes represent the error in percent and vertical axes probability, normalized to one. (\textbf{b}) Similar, for TLS and ROB.
\label{fig:reg_outlier}}
\end{figure}

Finally, the regression analysis was carried out for every dataset, and for each choice of the $\upbeta_i$, the obtained estimates $\hat{\upbeta}_i$ were defined as the average over the $10$ data realizations. Next, histograms were created based on these averages for the estimated coefficients, specifically the normalized histograms of the relative difference $(\upbeta_i - \hat{\upbeta}_i)/\upbeta_i$ ($i=0,\ldots,3$), expressed as a percentage, between the true value $\upbeta_i$ and the estimated value $\hat{\upbeta}_i$ of each regression parameter. The histograms of these percentage errors are shown in Figure~\ref{fig:reg_outlier}. In order to avoid a cluttered figure, the results of OLS, MAP and GLS are plotted in one panel and those of TLS and ROB in another.

From these histograms, it is clear that, for each parameter, GLS performs much better than OLS, MAP and TLS, with the latter failing completely. In case of GLS, the vast majority of relative errors is of the order of a few percent and certainly smaller than $20$\%. Overall, the most difficult to estimate parameter turns out to be $\upbeta_1$, which is associated with $\bar{n}_\mathrm{e}$. The robust estimation technique in MATLAB also delivers good results (in fact, not much worse than GLS), as it is designed to cope with outliers. However, we will see that in the next experiment ROB does not perform well at all.
%

\subsection{Effect of Logarithmic Transformation}

We next tested the effect of a logarithmic transformation, which is often used to transform a power-law regression model into a linear form. However, the logarithm alters the data distribution, which may lead to misguided inferences from OLS~\cite{mcdonald:stat_model,xiao:log}. Therefore, the flexibility offered by GLS is expected to be beneficial in this case, as it allows the observed distribution to deviate from the modeled distribution.

\subsubsection{Single Predictor Variable}

Again, we first performed a simple regression experiment involving a single predictor variable, with a power law deterministic model and additive Gaussian noise on all variables. In accordance with the typical situation of fitting fusion scaling laws to multi-machine data, the noise standard deviation was taken proportional to the simulated measurements, corresponding to a given set of relative error bars. \mbox{As a result}, in the logarithmic space the distributions were only approximately Gaussian, with the standard deviation given by the constant relative error on the original measurement (homoscedasticity). Ten points were chosen with predictor values $\upxi_n$ unevenly spread between zero and $60$. A power law was proposed to relate the unobserved $\upxi_n$ and $\upeta_n$:
\begin{equation*}
 \upeta_n = \upbeta_0\, \upxi_n^{\upbeta_1}, \quad n = 1,\ldots,10.
\end{equation*}

Then, Gaussian noise was added to the $\upxi_n$ and $\upeta_n$, corresponding to a substantial relative error of 40\%. We finally took the natural logarithm of all observed values $x_n$ and $y_n$, enabling application of the same linear regression methods that were used in the previous experiment. In this particular experiment, we chose $\upbeta_0 = 0.8$ and $\upbeta_1 = 1.4$, but we found that other values yield similar conclusions. Again, \mbox{$100$ data} replications were generated, allowing calculation of Monte Carlo averages.

The averages and standard deviations over all $100$ runs are given in Table~\ref{tab:reg_log_prop_stat}. Again, the results show that GLS is robust against the flawed model assumptions, now performing similar to TLS.
\begin{table}[H]
 \caption{Monte Carlo estimates of the mean and standard deviation for the parameters in a log-linear regression experiment with proportional additive noise on both variables.\label{tab:reg_log_prop_stat}}
 \centering
 \footnotesize
 \begin{tabular}{ccccccc}	
 	\toprule
 	 \textbf{Parameter} & \textbf{Original }& \textbf{GLS}    & \textbf{OLS}    & \textbf{MAP }   & \textbf{TLS}    & \textbf{ROB} \\
  	\midrule
 	 $\upbeta_0$ & 0.80 & \textbf{0.94} $\bm{\pm}$ \textbf{0.47} & 2.2 $\pm$ 2.3 & 3.0 $\pm$ 1.7 & 0.99 $\pm$ 0.70 & 2.72 $\pm$ 0.77\\
 	 $\upbeta_1$ & 1.40 & \textbf{1.39} $\bm{\pm}$ \textbf{0.11} & 1.19 $\pm$ 0.16 & 1.08 $\pm$ 0.26 & 1.41 $\pm$ 0.14 & 1.17 $\pm$ 0.11\\
 	\bottomrule
 \end{tabular}
\end{table}

\subsubsection{Multiple Predictor Variables}

In the last experiment with synthetic data, we studied the effect of a logarithmic transformation in a similar problem as the one described in Section~\ref{sec:outlier_mult}, but in the case of a power law. In particular, the variable $\upeta$ was calculated for the same range of values of the parameters $\upbeta_i$ as given in Equation \eqref{eq:range_beta}, but now according to a power law:
\begin{equation*} 
 \upeta = \upbeta_0\bar{n}_\mathrm{e}^{\upbeta_1} B_\mathrm{t}^{\upbeta_2} S^{\upbeta_3}.
\end{equation*}

Then, Gaussian noise was added to all variables. However, when applying the relatively low noise levels used in Section~\ref{sec:outlier_mult}, no significant differences were observed in the performance of GLS and MAP. Therefore, the noise levels for the predictor variables were augmented to $20$\% for $\bar{n}_\mathrm{e}$ (variable $x_1$), $5$\% for $B_\mathrm{t}$ (variable $x_2$) and $15$\% for $S$ (variable $x_3$). The level for $P_\mathrm{thr}$ was kept at $15\%$, as before. This~is still well within the maximum variability range that can be expected for the predictor variables in the ITPA database, as discussed in Section~\ref{sec:lh}.
\begin{figure}[H]
\centering
\includegraphics[width=0.54\textwidth]{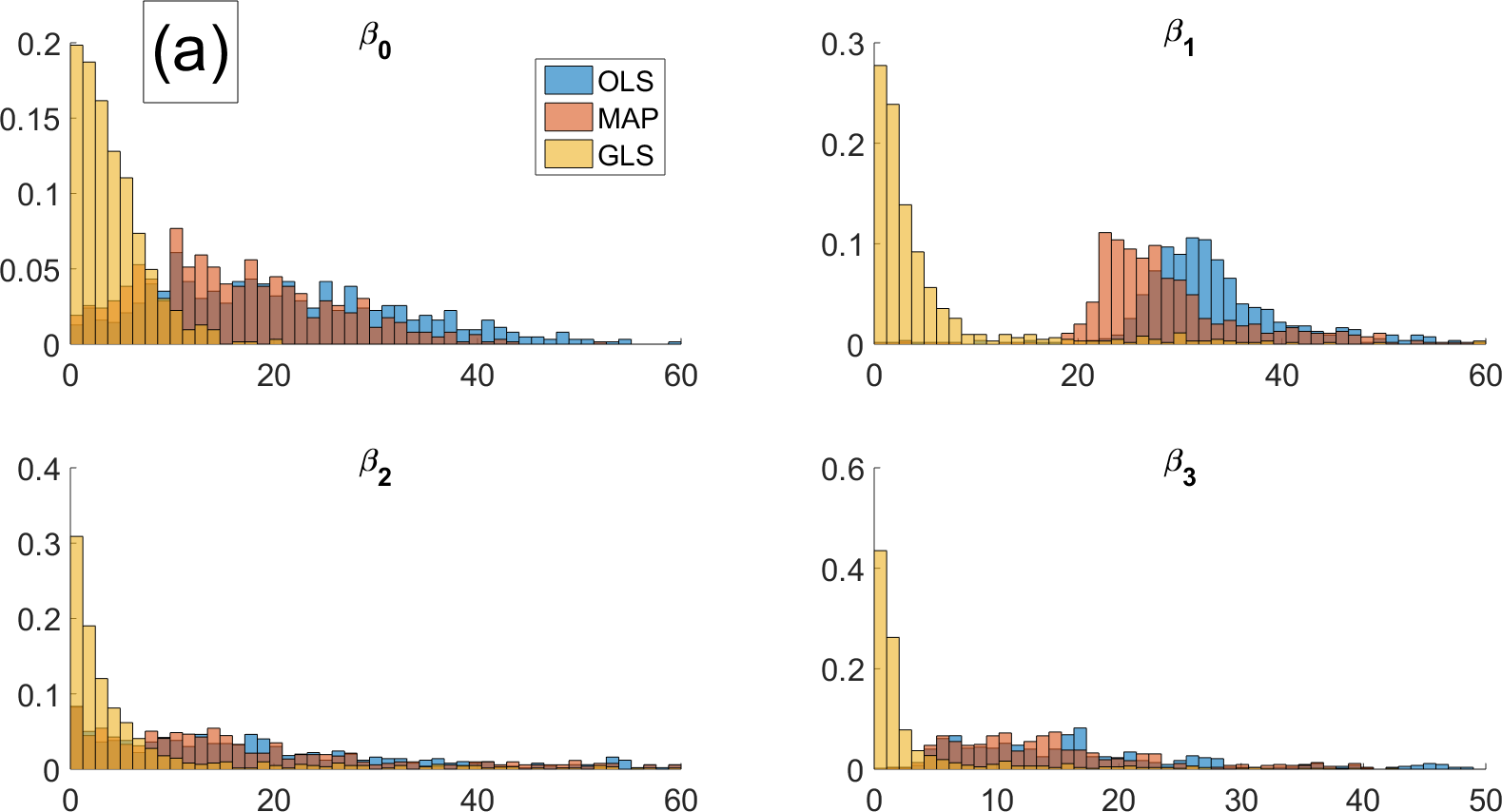}
\includegraphics[width=0.54\textwidth]{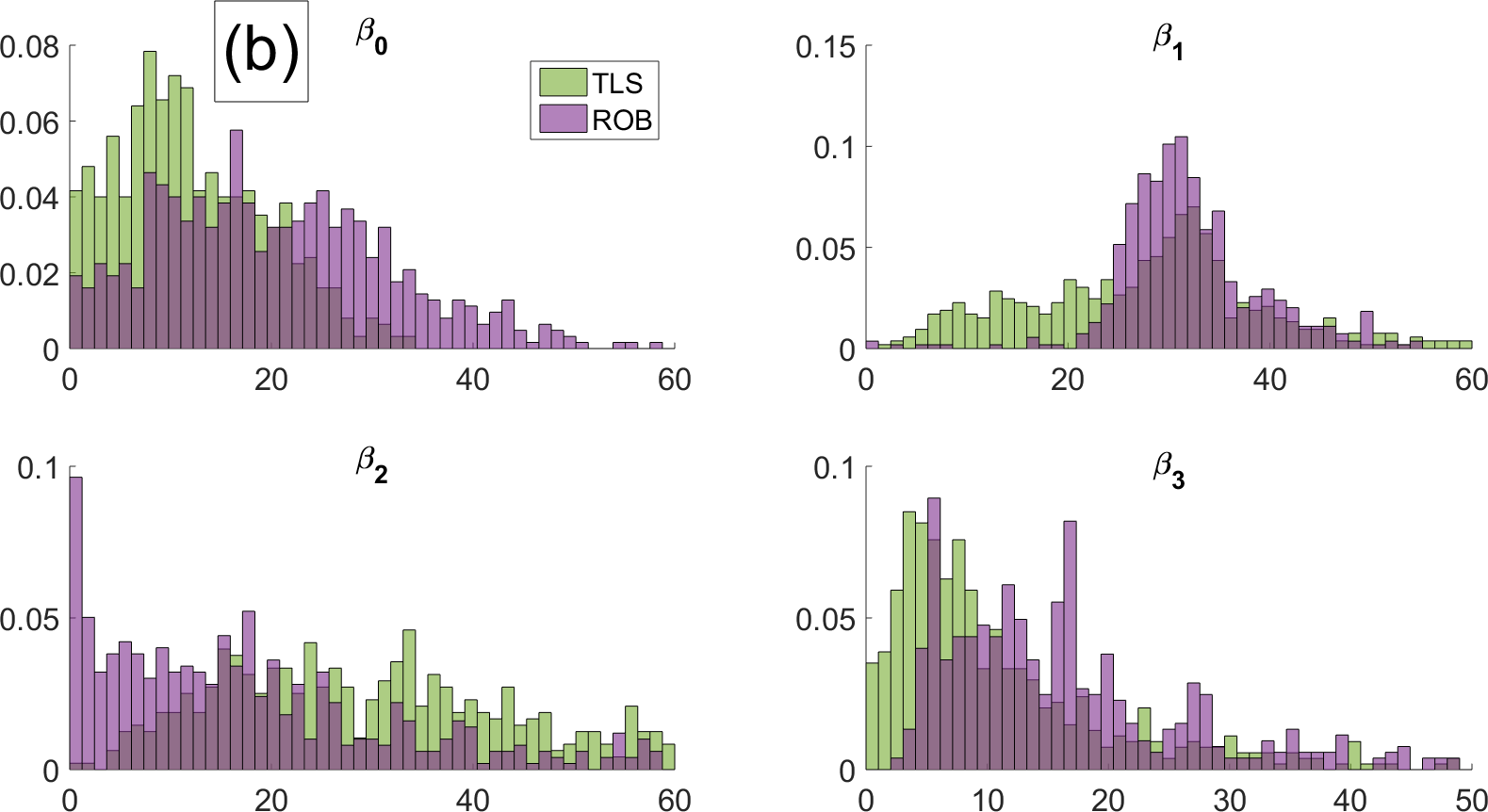}
\caption{Histograms of the relative error in estimating the regression coefficients $\upbeta_i$ by means of OLS, MAP and GLS for a power-law regression problem after a logarithmic transformation. Horizontal axes represent the error in percent and vertical axes probability, normalized to one. (\textbf{b}) Similar, for TLS and ROB.\label{fig:reg_log}}
\end{figure}

After adding the noise, all data were transformed to the logarithmic domain, and 10 datasets were generated for each combination of regression coefficients. Subsequently, linear regression analysis was applied to each of the log-transformed datasets. The coefficient estimates, defined as the average over the 10 replications, were then compared among the various regression methods, as shown in Figure~\ref{fig:reg_log}. Again, the normalized histograms of the relative error on the estimated parameters are displayed, showing the consistently better performance of GLS over all other methods tested, including TLS and ROB. For~GLS, the errors on $\upbeta_0$ and $\upbeta_1$ are the largest, compared to those on $\upbeta_2$ and $\upbeta_3$, but the majority is still below 20\%. As for $\upbeta_0$, the slightly inferior performance of GLS relative to the results with outliers in Section~\ref{sec:outlier_mult} is simply due to the fact that $\log\upbeta_0$ for the lowest values of $\upbeta_0$ is negligibly small compared to $\log\upeta - \log\upbeta_0$.

\section{Power Threshold Scaling} \label{sec:pow}

We finally come to the application of power threshold scaling using real-world data from the ITPA database for all variables, including the response variable $P_\mathrm{thr}$. We start with log-linear regression and then apply nonlinear regression analysis. Next, we perform a simple analysis of the influence of the error bars on the estimation results, and we finally provide a discussion of the results in this section.

\subsection{Linear Scaling} \label{sec:pow_lin}

We first followed the standard practice of transforming to the logarithmic scale to estimate the coefficients $\upbeta_0$, $\upbeta_1$, $\upbeta_2$ and $\upbeta_3$ in Equation \eqref{eq:power} via linear regression. In the GLS method, we introduced additional parameters $\upsigma_{\up{obs},\alpha}$ ($\alpha=1,\ldots,N_\up{t}$), one for each of the $N_\up{t}=8$ tokamaks contributing data to the scaling. That is, if a certain data point with index $n$ originated from tokamak $\alpha$, then in term $n$ of the objective function in~Equation \eqref{eq:obj_fun}, an observed distribution was used, parameterized by means of the $\upsigma_{\up{obs},\alpha}$ corresponding to that machine. The $\upsigma_{\up{obs},\alpha}$ serve a similar purpose as the parameter $\upsigma_\up{obs}$ defined above, except that they describe the observed standard deviations of the {logarithmic} power threshold. This, of course, corresponds to the relative errors on the power threshold itself. To calculate $\upsigma_\mathrm{mod}$ for each data point, we used the relative measurement error bars quoted in the database (typically $4$\% for $\bar{n}_\mathrm{e}$, $1$\% for $B_\mathrm{t}$, $3$\% for $S$ and $15\%$ for $P_\mathrm{thr}$). Considering the discussion in Section~\ref{sec:lh} regarding other sources of uncertainty, it is clear that the $\upsigma_{\up{obs},\alpha}$ will need to take into account other, ``unexpected'' uncertainty sources, hence increasing the flexibility of the method.

In this analysis, we compared GLS only with OLS and the powerful MAP method. The results on the IAEA02 data are given in Table~\ref{tab:res_power_log_02}. The predictions for ITER are also shown, for two typical densities ($0.5$ and $1.0\times 10^{20}$ $\up{m}^{-3}$). All estimates are accompanied by their 95\% credible intervals obtained from 100~bootstrap samples (artificial datasets). We stress that this notion of a credible interval corresponds to the standard Bayesian definition of an interval wherein the true value of a stochastic variable is assumed to lie with a certain probability (e.g., $0.95$).


The estimates by GLS of the parameters $\upsigma_{\up{obs},\upalpha}$ (observed standard deviation on $\log P_\up{thr}$), for each of the devices contributing to the IAEA02 data, were expressed as a relative error on the bootstrap-averaged $P_\up{thr}$. These relative errors and their credible intervals are given in Table~\ref{tab:res_sigma_log_02}. The relative error on the power threshold lies around 15\% to 30\% for the various machines, except for ASDEX, where the uncertainty reaches a higher level of about 40\%. On average, this yields an estimated error of $24.2$\% for $P_\mathrm{thr}$, which is quite somewhat higher than the average of $15\%$ mentioned in the database, although still considerably lower than the upper bound of $38\%$, as calculated in Section~\ref{sec:lh}. Again, this is an indication of additional sources of uncertainty, on top of mere measurement error, causing the data points to deviate from the proposed regression model, as discussed already in Section~\ref{sec:lh}. That extra uncertainty is captured by the GLS method.

\begin{table}[H]
 \caption{Estimates of regression parameters and predictions for ITER in log-transformed linear scaling of the H-mode threshold power using the IAEA02 dataset. The bootstrap averages are given, as well as the 95\% credible intervals (CI).\label{tab:res_power_log_02}}
 \centering
 \small
 \begin{tabular}{@{}cccccccc}	
 	\toprule
  \textbf{Method }    &   & $\hat{\bm\upbeta}_\textbf{0}$ & $\hat{\bm\upbeta}_\textbf{1}$ & $\hat{\bm\upbeta}_\textbf{2}$ & $\hat{\bm\upbeta}_\textbf{3}$ & $\hat{\textbf{P}}_{\up{\textbf{thr}},\textbf{0.5}}$ (\textbf{MW
	}) & $\hat{\textbf{P}}_{\up{\textbf{thr}},\textbf{1.0}}$ (\textbf{MW}) \\
  \midrule
  \multirow{2}{*}{OLS} & Average & 0.0507 & 0.485 & 0.873 & 0.843 & 38.0 & 53.2 \\
         & CI  & $\pm 0.0060$ & $\pm 0.073$ & $\pm 0.061$ & $\pm 0.041$ & $\pm 4.4$ & $\pm 8.0$ \\
  \midrule
  \multirow{2}{*}{MAP} & Average & 0.0449 & 0.567 & 0.867 & 0.901 & 45.6 & 67.6 \\
         & CI  & $\pm 0.0051$ & $\pm 0.078$ & $\pm 0.069$ & $\pm 0.039$ & $\pm 5.0$ & $\pm 9.6$ \\
  \midrule
  \multirow{2}{*}{GLS} & Average & 0.0426 & 0.660 & 0.795 & 0.946 & 48.3 & 76.4 \\
         & CI  & $\pm 0.0042$ & $\pm 0.069$ & $\pm 0.059$ & $\pm 0.034$ & $\pm 4.7$ & $\pm 9.8$ \\
 	\bottomrule
 \end{tabular}
\end{table}

\begin{table}[H]
 \caption{Estimates of the observed standard deviations $\upsigma_{\up{obs},\upalpha}$ of the logarithmic power threshold, expressed as percentage errors on $P_\up{thr}$ itself, for the tokamaks contributing to the IAEA02 dataset, obtained using log-transformed linear scaling. The bootstrap averages are given, as well as the 95\% credible intervals (CI).\label{tab:res_sigma_log_02}}
 \centering
 \small
 \begin{tabular}{@{}ccccccccc}	
 	\toprule
      & \textbf{ASDEX
			} & \textbf{AUG
			} &\textbf{ CMOD
			} & \textbf{DIII-D
			} & \textbf{JET
			} & \textbf{JFT-2M
			} & \textbf{JT-60U
			} & \textbf{PBXM
			} \\
  \midrule
  Average (\%) & 41.8 & 23.0 & 22.0 & 15.7 & 24.6 & 15.9 & 22.8 & 27.6 \\
  CI (\%)  & $\pm 5.3$ & $\pm 1.4$ & $\pm 1.1$ & $\pm 1.8$ & $\pm 2.0$ & $\pm 1.2$ & $\pm 2.3$ & $\pm 2.9$ \\
  \bottomrule
 \end{tabular}
\end{table}

\subsection{Nonlinear Scaling} \label{sec:pow_nlin}

Next, we show the results of nonlinear regression in the original data space, \emph{i.e.}, without logarithmic transformation. Whereas this prevents an analytic solution using OLS, the advantage is that the distribution of the data is left undistorted~\cite{mcdonald:stat_model,xiao:log}, while the implementation of OLS, MAP and GLS is not significantly more complex. Indeed, the distribution of the right-hand side in Equation \eqref{eq:power} can be approximated by a Gaussian with mean $\upmu_\mathrm{mod}=\upbeta_0\, \bar{n}_\mathrm{e}^{\upbeta_1} B_\mathrm{t}^{\upbeta_2} S^{\upbeta_3}$ and standard deviation $\upsigma_\mathrm{mod}$, given by:
\begin{equation*}
 \upsigma_\mathrm{mod}^2 = \upsigma_{P_\mathrm{thr}}^2 +\upmu_\mathrm{mod}^2\left[\upbeta_1^2\left(\frac{\upsigma_{\bar{n}_\mathrm{e}}}{\bar{n}_\mathrm{e}}\right)^2 + \upbeta_2^2\left(\frac{\upsigma_{B_\mathrm{t}}}{B_\mathrm{t}}\right)^2 + \upbeta_3^2\left(\frac{\upsigma_{S}}{S}\right)^2\right].
\end{equation*}

Hence, the modeled standard deviations depend on the measurements (heteroscedasticity). Nevertheless, in defining the observed standard deviations $\upsigma_{\up{obs},\alpha}$, we introduced an approximation assuming constant error bars for all measurements from a single machine. This assumption may be relaxed in the future.

The results of the scalings and predictions are presented in Tables~\ref{tab:res_power_pow_02} and~\ref{tab:res_sigma_pow_02}. We compared GLS with OLS and MAP using uniform priors. It may be possible to derive even less informative priors for MAP, as was done in the log-linear case in Section~\ref{sec:pow_lin} (and see~\cite{preuss:all_variables,Von_toussaint:fitting}), but this was not pursued here. Moreover, even in the log-linear analysis, we observed only a marginal difference between the results under various choices of priors.

\begin{table}[H]
 \caption{Estimates of regression parameters and predictions for ITER in power-law scaling on the original scale of the H-mode threshold power using the IAEA02 dataset. The bootstrap averages are given, as well as the 95\% credible intervals (CI).\label{tab:res_power_pow_02}}
 \centering
 \small
 \begin{tabular}{@{}cccccccc}	
 	\toprule
  \textbf{Method }    &   & $\hat{\bm\upbeta}_\textbf{0}$ & $\hat{\bm\upbeta}_\textbf{1}$ & $\hat{\bm\upbeta}_\textbf{2}$ & $\hat{\bm\upbeta}_\textbf{3}$ & $\hat{\textbf{P}}_{\up{\textbf{thr}},\textbf{0.5}}$ (\textbf{MW}) & $\hat{\textbf{P}}_{\up{\textbf{thr}},\textbf{1.0}}$ (\textbf{MW}) \\
  \midrule
  \multirow{2}{*}{OLS} & Average & 0.0274 & 0.773 & 0.96 & 1.038 & 69 & 118 \\
         & CI  & $\pm 0.0083$ & $\pm 0.090$ & $\pm 0.10$ & $\pm 0.071$ & $\pm 15$ & $\pm 32$ \\
  \midrule
  \multirow{2}{*}{MAP} & Average & 0.0425 & 0.643 & 0.788 & 0.933 & 44.2 & 69.1 \\
         & CI  & $\pm 0.0041$ & $\pm 0.074$ & $\pm 0.079$ & $\pm 0.034$ & $\pm 3.8$ & $\pm 8.2$ \\
  \midrule
  \multirow{2}{*}{GLS} & Average & 0.0397 & 0.715 & 0.751 & 0.984 & 51.6 & 84.7 \\
         & CI  & $\pm 0.0036$ & $\pm 0.071$ & $\pm 0.081$ & $\pm 0.031$ & $\pm 4.0$ & $\pm 8.8$ \\
 	\bottomrule
 \end{tabular}
\end{table}
\begin{table}[H]
 \caption{Estimates of the observed standard deviations $\upsigma_{\up{obs},\alpha}$ of the power threshold $P_\up{thr}$, expressed as percentage errors, for the machines contributing to the IAEA02 dataset, obtained using power-law scaling. The bootstrap averages are given, as well as the 95\% credible intervals (CI).\label{tab:res_sigma_pow_02}}
 \centering
 \small
 \begin{tabular}{@{}ccccccccc}	
 	\toprule
 	& \textbf{ASDEX} &\textbf{ AUG} & \textbf{CMOD} &\textbf{DIII-D }&\textbf{ JET} & \textbf{JFT-2M} & \textbf{JT-60U} & \textbf{PBXM} \\
  \midrule
  Average (\%) & 35.8 & 21.2 & 20.4 & 15.9 & 22.4 & 15.7 & 22.3 & 27.7 \\
  CI (\%) & $\pm 9.1$ & $\pm 4.3$ & $\pm 3.4$ & $\pm 2.4$ & $\pm 3.8$ & $\pm 2.2$ & $\pm 4.6$ & $\pm 8.1$ \\
  \bottomrule
 \end{tabular}
\end{table}

It should also be mentioned that, in obtaining Table~\ref{tab:res_sigma_pow_02}, we again calculated relative errors from the observed standard deviations estimated by GLS. However, this time, the relative errors are not the same for all measurements coming from a single machine, so we calculated an average for each machine (and similar for the credible interval). The resulting errors on $P_\mathrm{thr}$ are relatively similar to those using log-linear scaling, with an average over all devices of 22.7$\%$, which is again higher than the $15\%$ expected from measurement error only.

\subsection{Influence of Error Bars}

In the last couple of experiments, we intended to assess the sensitivity of the regression analysis on the accuracy of the error bars on the ITPA data. A systematic study of this influence is outside the scope of this paper, and as a first simple test, we doubled the error bars on all root ITPA variables (basically the electron density and the magnetic field together with various geometrical plasma parameters and sources of input power), which were used for calculation of the variables involved in the power threshold scaling law. On average, over all machines, this resulted in the following derived error bars: $9\%$ on $\bar{n}_\mathrm{e}$, $2\%$ on $B_\mathrm{t}$, $5\%$ on $S$ and $32\%$ on $P_\mathrm{thr}$. Again, these are all below the maxima quoted in Section~\ref{sec:lh}.

We then performed power-law regression with MAP and GLS on the ITPA data using these larger error bars; the results are given in Table~\ref{tab:res_power_pow_eb2_02} \cite{footnote4}. It is observed that, based on MAP, the predictions for ITER are lowered relative to the analysis with the original error bars in Section~\ref{sec:pow_nlin}. In contrast, the predictions by GLS remain about the same as before. On the other hand, the GLS estimates of the observed standard deviations, listed in Table~\ref{tab:res_sigma_pow_eb2_02}, are increased for all devices. This is how GLS accommodates the increased error bars \mbox{on the data}.

\begin{table}[H]
 \caption{Estimates of regression parameters and predictions for ITER in power-law scaling on the original scale of the H-mode threshold power using the IAEA02 dataset with all error bars (on the root quantities) doubled.\label{tab:res_power_pow_eb2_02}}
 \centering
 \small
 \begin{tabular}{@{}lcccccc}	
 	\toprule
  \textbf{Method }    & $\hat{\bm\upbeta}_\textbf{0}$ & $\hat{\bm\upbeta}_\textbf{1}$ & $\hat{\bm\upbeta}_\textbf{2}$ & $\hat{\bm\upbeta}_\textbf{3}$ & $\hat{\textbf{P}}_{\up{\textbf{thr}},\textbf{0.5}}$ (\textbf{MW}) & $\hat{\textbf{P}}_{\up{\textbf{thr}},\textbf{1.0}}$ (\textbf{MW}) \\
  \midrule
  MAP & 0.0436 & 0.581 & 0.828 & 0.900 & 41.0 & 61.3 \\
  GLS & 0.0393 & 0.725 & 0.742 & 0.990 & 52.1 & 86.2 \\
  \bottomrule
 \end{tabular}
\end{table}
\begin{table}[H]
 \caption{Estimates of the observed standard deviations $\upsigma_{\up{obs},\upalpha}$ of the power threshold $P_\up{thr}$, expressed as percentage errors, for the machines contributing to the IAEA02 dataset with all error bars doubled, obtained using power-law scaling.\label{tab:res_sigma_pow_eb2_02}}
 \centering
 \small
 \begin{tabular}{@{}cccccccc}	
 	\toprule
  \textbf{ASDEX} & \textbf{AUG} &\textbf{ CMOD} & \textbf{DIII-D} & \textbf{JET} & \textbf{JFT-2M} & \textbf{JT-60U} & \textbf{PBXM} \\
 \midrule
  49.5 & 35.9 & 31.7 & 24.9 & 32.9 & 27.6 & 38.9 & 47.7 \\
    \bottomrule
 \end{tabular}
\end{table}

In another simple test, we changed the error bars on $\bar{n}_\mathrm{e}$, $B_\mathrm{t}$, $S$ and $P_\mathrm{thr}$ to values computed from the {average} percentages mentioned earlier in Section~\ref{sec:lh}: $4$\% for $\bar{n}_\mathrm{e}$, $1$\% for $B_\mathrm{t}$, $3$\% for $S$ and $15\%$ for $P_\mathrm{thr}$. These are averages over all machines, rendering the final absolute error bars (standard deviations), computed from the relative errors, less precise. The estimation results using power-law regression with MAP and GLS are shown in Table~\ref{tab:res_power_pow_eb_av_02}. The results of both methods are clearly affected by the averaging step, but again, MAP is seen to be more sensitive to the change in the error bars compared to GLS, which maintains estimates in a similar range as those given in Tables~\ref{tab:res_power_log_02} and \ref{tab:res_power_pow_02}. The estimates of the observed standard deviations, given in Table~\ref{tab:res_sigma_pow_eb_av_02}, are adjusted accordingly by GLS.

\begin{table}[H]
 \caption{Estimates of regression parameters and predictions for ITER in power-law scaling on the original scale of the H-mode threshold power using the IAEA02 dataset with averaged error bars.\label{tab:res_power_pow_eb_av_02}}
 \centering
 \small
 \begin{tabular}{@{}lcccccc}	
 	\toprule
  \textbf{Method }    & $\hat{\bm\upbeta}_\textbf{0}$ & $\hat{\bm\upbeta}_\textbf{1}$ & $\hat{\bm\upbeta}_\textbf{2}$ & $\hat{\bm\upbeta}_\textbf{3}$ & $\hat{\textbf{P}}_{\up{\textbf{thr}},\textbf{0.5}}$ (\textbf{MW}) & $\hat{\textbf{P}}_{\up{\textbf{thr}},\textbf{1.0}}$ (\textbf{MW}) \\
  \midrule
  MAP & 0.0488 & 0.552 & 0.807 & 0.862 & 35.1 & 51.5 \\
  GLS & 0.0429 & 0.647 & 0.780 & 0.938 & 45.7 & 71.5 \\
  \bottomrule
 \end{tabular}
\end{table}
\begin{table}[H]
 \caption{Estimates of the observed standard deviations $\upsigma_{\up{obs},\alpha}$ of the power threshold $P_\up{thr}$, expressed as percentage errors, for the machines contributing to the IAEA02 dataset with averaged error bars, obtained using power-law scaling.\label{tab:res_sigma_pow_eb_av_02}}
 \centering
 \small
 \begin{tabular}{@{}cccccccc}	
 	\toprule
  \textbf{ASDEX} & \textbf{AUG} & \textbf{CMOD} & \textbf{DIII-D} & \textbf{JET} & \textbf{JFT-2M} & \textbf{JT-60U} & \textbf{PBXM} \\
  \midrule
  29.6 & 19.1 & 20.5 & 19.5 & 22.5 & 18.1 & 18.7 & 20.4 \\
   \bottomrule
 \end{tabular}
\end{table}

\subsection{Discussion}

Several interesting observations can be made from the experiments regarding the power threshold scaling in this section. First, considering Tables~\ref{tab:res_power_log_02} and \ref{tab:res_power_pow_02}, it should be noted that there are several instances where the regression parameters estimated by OLS differ significantly from those obtained by GLS. For~log-linear regression, this is particularly the case for the dependence of the power threshold on density and surface area and for the predicted power thresholds for ITER, as shown by the non-overlapping credible intervals. For power-law regression, the difference is rather situated in the dependence on the magnetic field. In this case, the power thresholds predicted by OLS are also quite different from the results given by GLS, but this time, the credible intervals on the OLS estimates are so wide, that they overlap with those obtained from GLS. Apart from this discrepancy, the three methods provide comparable absolute error bars on their estimates.

Furthermore, we see that the correspondence between GLS and MAP is significantly better, although the remaining differences become particularly clear for the predicted power at higher density in ITER. The estimate by GLS is higher than that provided by MAP, especially for power-law scaling.

In addition, and quite remarkably, when comparing the coefficient estimates and predictions obtained by GLS between the linear and nonlinear case, relatively consistent results can be noted. The same goes for the MAP estimates. In contrast, OLS provides widely different results, depending on whether a linear (log-transformed) or nonlinear (power-law) model is used. The relatively good consistency of the GLS estimates across regression models is a solid argument in favor of the method.

Another noteworthy point comes from the results of the two additional tests with increased and averaged error bars. They indicate that for MAP (and maximum likelihood) regression, reliable estimates of the variability of the measurements is important. However, as discussed in Section~\ref{sec:lh}, the standard error bars that were used in the analysis in Sections~\ref{sec:pow_lin} and \ref{sec:pow_nlin} are small compared to the actual variability of the data around the theoretical scaling law. Hence, one could speculate whether the results of the MAP analysis are in fact trustworthy, given its sensitivity to the error bars on the data. Therefore, at least for MAP, it would be better to encode the available information on the error bars in sufficiently wide prior distributions (which, incidentally, would be possible for GLS, too).

A related comment is that GLS is clearly less vulnerable to inaccurate error specification compared to MAP. The mechanism behind this behavior is similar to the one that makes GLS less sensitive to outliers, \emph{i.e.}, the observed standard deviation is able to capture deviations from the expected data variability with respect to the model. In the simple implementation of the GLS method used in the present paper, the distinction that is made between the modeled and observed standard deviation is the main difference between GLS and MAP.

\section{Conclusions} \label{sec:concl}

Regression and scaling laws represent crucial tools in science in general and in the analysis of complex physical systems in particular. We have presented geodesic least squares regression (GLS) as a method that is able to handle large uncertainties on the data and on the regression model, and we have demonstrated its application to power-law regression. Operating on a manifold of probability distributions, GLS has the advantage that its results can be easily visualized in the case of the univariate Gaussian distribution. However, GLS is sufficiently flexible to allow tackling much more general regression problems within the same framework.

We have shown two examples of the enhanced robustness of the method using synthetic data. GLS showed a better stability in the presence of outliers and under a logarithmic transformation of a power-law, compared to established techniques. In addition, we have addressed the scaling of the L-H power threshold in magnetically-confined fusion plasmas. On the basis of data from a multi-machine database, it was observed that geodesic least squares provides estimates of regression parameters and predictions that are consistent across different regression models, in contrast to ordinary least squares. Furthermore, because GLS allows the data uncertainty predicted by the model to be different from the empirically observed uncertainty, whereas with maximum \emph{a posteriori} they are, by design, the same, GLS is more flexible and robust at the same time. As a consequence, the degrees of freedom provided by the parameters of the regression model better serve their actual purpose: to parameterize a model that best describes a trend in the data, with minimal distraction by the data ``noise''.

In future work, we intend to present a more general formulation of geodesic least squares, targeted at a wider class of regression problems. In addition, various theoretical performance issues need to be addressed, including uniqueness and convergence properties of the optimization problem, asymptotic behavior of the parameter estimates, \emph{etc}. On the practical side, we aim at establishing a broader basis for the performance of the GLS method on simulated data. This should increase the confidence over a wider range of regression problems, as well as deviations from the regression model.

Finally, although we have noted that GLS performs regression on a probabilistic manifold, we have actually made little use of the geometrical structure of the manifold, save for calculating geodesic distances. Nowadays, there are various schemes, more sophisticated than a least-squares approach, to perform regression on manifold-valued data. From that point of view, one can expect advantages of a method performing regression between probability distributions, each of them containing more information than structureless data points in a Euclidean space. One possibility that we will explore in future work is a Bayesian regression method on a probabilistic manifold, by describing the distribution corresponding to the regression model intrinsically on the manifold~\cite{pennec:jmiv}. At the same time, this will provide uncertainty estimates on the parameters through the posterior distribution.
\acknowledgments{Acknowledgments}
The author wishes to acknowledge the ITPA Topical Groups on Transport and Confinement and on Pedestal and Edge Physics for maintaining and kindly providing the data in the H-mode~threshold~databases.

\conflictofinterests{Conflicts of Interest}
The author declares no conflict of interest.


\bibliographystyle{mdpi}

\begin{thebibliography}{----}
\bibitem {footnote1}
We follow standard notational practice from differential geometry with respect to index placement in the following definitions for the metric, Christoffel symbols and geodesic distance. However, in the remainder of the paper we will revert to subscript indices only, in order to avoid other notational problems.
\bibitem {footnote2}
 Whereas the most recent update of the database dates from 2008~\cite{martin:power08}, we used the earlier version from 2002, because it allows a better illustration of the advantages of GLS with respect to other methods. The reason is that the data in the most recent version is significantly better conditioned, in which case even a simple regression technique such as OLS turns out to be able to provide acceptable estimates of the regression parameters. This point is not relevant for the present discussion, as here our aim is to demonstrate the advantages of GLS in cases where the data are not in the best shape.
\bibitem{footnote3}
 We use the notation $\upeta$ for the response variable instead of $P_\mathrm{thr}$ because in this experiment $\upeta$ is generated artificially and therefore it is not necessarily related to the actual power threshold in fusion devices.
\bibitem{footnote4}
OLS is not repeated here because it does not depend on the error bars.
\end{thebibliography}

%
\end{document}